# How to Draw Graphs:
## Seeing and Redrafting Large Networks in Security and Biology

Bob Blakley[1], GR Blakley[2], Sean Blakley[3]

[1]Citigroup
bob.blakley@citi.com
[2]Mathematics Department, Texas A&M University
grb@ieee.org
[3]Physics Department, Texas A&M University
SMB784@tamu.edu

**Abstract.** A graph is a mathematical object consisting of a set of vertices and a set of edges connecting vertices. Graphs can be drawn on paper in various ways, but until recently all published methods of drawing graphs have had undesirable properties: (i) for graphs which are not plane embeddable, intersections between the lines representing edges appear at points which are not vertices, creating the appearance of vertices where none exist, (ii) vertex labels can be placed inside vertex symbols, but there is no consistent, logical, and visually clean place to put edge labels, and (iii) representations of large graphs are visually dense and difficult to interpret. This paper describes a new "cartographic" method of drawing graphs which solves all of these problems, and has other advantages as well. Complements, comparisons and contrasts of graphs are usually better shown cartographically than in node-link form.

## 1 Introduction

Centuries ago people started using node-link diagrams to help them contemplate problems and explain discoveries in the theory of graphs, an abstract axiomatic science with no pictorial content of its own. Such diagrams represent the $v$ vertices (or "nodes") of a graph $G$ by dots or small circles, and represent $G$'s $e$ edges (or "links") by curves or line segments. Node-link diagrams were and still are helpful in dealing with some small or sparse graphs. But the vast majority of big or dense [GY04, p. 57] graphs (e.g. graphs such that $10^6 < v^3 < e^2$) have node-link representations which are useless hairballs.

The deluge of big dense graphs arising from recent studies in biology, computer security, and information science has led several investigators to rethink graph visualization and employ cartographic diagrams to guide and explain their work on simple graphs, multigraphs, graphs with loops, digraphs, and networks. Such diagrams use latitudes to represent vertices and longitudes to represent edges. Consequently they cannot be cluttered, crowded, or ambiguous. No cartographic representation of a graph can be a hairball.

This paper treats cartographic representations of graphs and related objects, as well as some developments in graph theory itself which they have stimulated. Geometry's synthetic side was already very rich when its coordinatized analytic side first appeared. Analogously, graph theory's many synthetic resources currently make scant reference to coordinates. But the influence of cartographic visualization seems at present to entail coordinatization, and thus to lead naturally to an analytic graph theory.

## 2   History of Cartographic Representations of Graphs

Edward Tufte has identified a good canvas to draw graphs on. It is a $1000 tatami screen — a 1-meter-high by 2-meter-wide video screen with resolution finer than 1 pixel per millimeter. However, a node-link drawing of even a small Kuratowski complete graph [WE96, pp. 8, 473] $K25$, which has 25 vertices and 300 edges is a crowded, cluttered, unintelligible hairball [MA13] on such a screen.

On the other hand, a cartographic drawing of the much larger 63-vertex, 1953-edge Kuratowski complete $K63$ graph can easily and intelligibly be drawn on the bottom 7% of the same screen. Indeed, uncrowded, uncluttered, unambiguous, intelligible cartographic representations of fifteen graphs — each with 63 vertices and any possible number of edges — can be displayed simultaneously in a vertical stack on such a screen.

Tufte's 1983 book, "The Visual Display of Quantitative Information" [TU83], considered how to present large data sets in ways which conveyed their import quickly and efficiently. Shortly thereafter, one of the authors [YL87] introduced cartographic displays — using latitudes to represent vertices and longitudes to represent edges — to draw simple graphs arising from the drafting of flowcharts. In 1999, Andrew McAllister [MC99] drew a cartographic representation of a multigraph in a treatment of a linear arrangement problem. And in 2012, William Longabaugh [LO12] introduced the BioFabric software, developed at the Institute for Systems Biology, and demonstrated its use in the display of cartographic representations of large networks.

## 3   How Not to Draw a Graph

The node-link graph representation consists of a collection of dots or circles representing the graph's vertices, arranged on the page in some pattern which makes sense to whoever is doing the drawing, connected by curves or line segments representing the graph's edges. For a small graph $G$, the result looks like Figure 1:

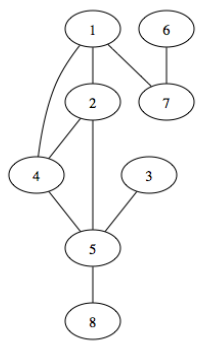

**Fig. 1.** A small graph $G$ drawn in node-link representation

Representing a large graph in this way creates a drawing which does little to aid human understanding, as Figure 2 (a random directed graph $H$ of order 24 with loops whose vertices have average degree 12) illustrates:

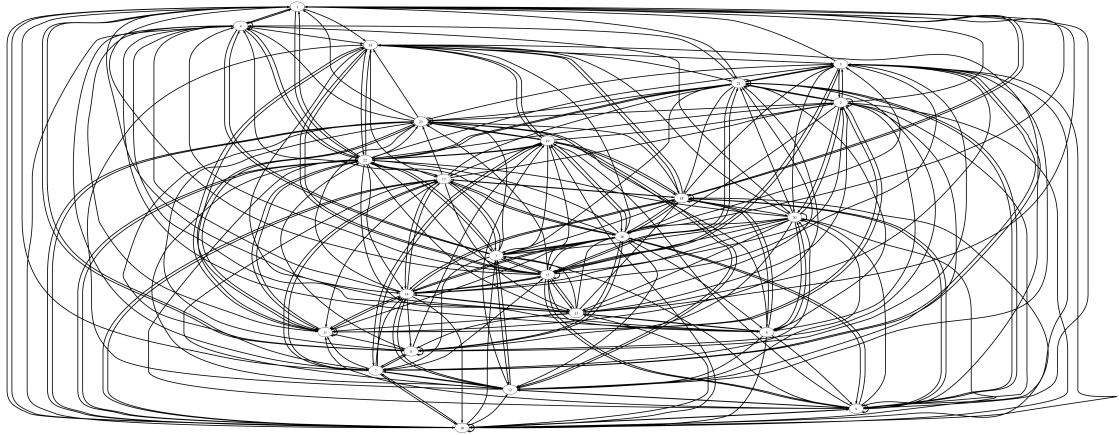

**Fig. 2.** A large digraph $H$ drawn as a node-link hairball

Ma and Muelder [MA13] and Longabaugh [LO12] call this representation a "hairball". The hairball representation has several shortcomings:

1. The representation is not canonical; vertices and edges do not occupy regular, predictable locations.
2. There is an obvious place to put vertex labels (inside the circles representing the vertices), but there is no obvious place to put edge labels.
3. Lines representing edges intersect at points which are not vertices.
4. The drawing is visually dense: it is difficult to tell whether any given pair of vertices is connected by an edge, it is difficult to tell which vertices any given edge connects, and it is difficult to determine the degree of each vertex.
5. Because line segments representing edges can cross at very acute angles, it is difficult to distinguish some pairs of edges.

Many alternative representations have been proposed to address these shortcomings. For example, it is possible to canonicalize the hairball representation by placing the vertex circles at equal intervals around the circumference of a circle, and drawing all the edges as straight lines connecting the circles representing edge's endpoint vertices. The result is still a hairball, and doesn't do anything to fix problems (2)-(5), as Figure 3 illustrates:

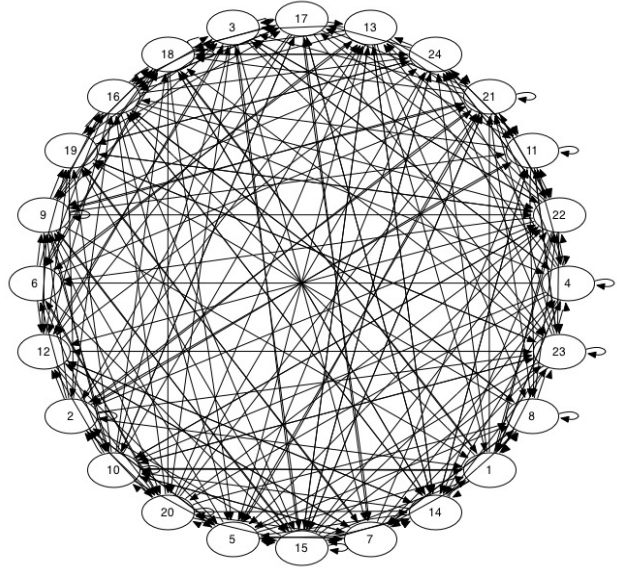

**Fig. 3.** The large digraph $H$ drawn as a canonicalized node-link hairball

## 4  Augmented Incidence Matrices

The adjacency matrix of a graph is a very economical description of that graph. There is, however, no obvious way to proceed from a graph's adjacency matrix to either a node-link drawing or a cartographic drawing.

Incidence matrices are another story. In fact, it is hardly an exaggeration to say that each of the very numerous incidence matrices of a graph $G$ amounts to a cartographic drawing of $G$, and that these cartographic drawings can be very disparate, and that many can be quite informative about numerous different properties of $G$. One of them may make it obvious that $G$ is bipartite, or is a chain, or is planar.

To facilitate drawing graphs, we define the augmented incidence matrix of a graph $G$ as follows:

*Let the graph $G$ of order $n$ be a subgraph of the Kuratowski complete graph $Kn$.*

*Define $D$ (the augmented incidence matrix of $G$) to be the incidence matrix of $Kn$, with columns representing edges which do not occur in $G$ replaced by columns in which each entry is 0.*

Clearly the augmented incidence matrix $D$ has exactly $a(n) = com(n, 2) = [n][n-1]/2$ columns. (Note that the augmented incidence matrix of a graph with loops will have $[n][n+1]/2$ columns, and the augmented incidence matrix of a digraph without loops will have $[n][n-1]$ columns).

# 5    How to Draw One Graph

The canonicalized hairball in Figure 3 fixes most of the node-link representation's problems with depicting vertices: it provides a canonical location for each vertex and a standard place to put each vertex's label. What's needed is a representation which preserves these virtues, but also provides a standard place to apply a label to each edge without overlaying the representation of the graph itself. The representation must also eliminate intersections of lines representing edges at locations which are not representations of vertices, and it must provide a visually comprehensible separation of edges from edges and vertices from vertices.

The cartographic representation satisfies these requirements. We draw the cartographic representation of the simple graph $G$ shown in Figure 1 as follows:

First, observe that the graph has 8 vertices. Start, therefore, with the incidence matrix for the complete graph $K8$. Now draw black vertical line segments connecting the "1" entries in each column of the augmented incidence matrix, and then replace the matrix with gray horizontal lines at the vertical center of each row and gray line segments at the horizontal center of each column. The result is a cartographic drawing of $K8$, as shown in Figure 4:

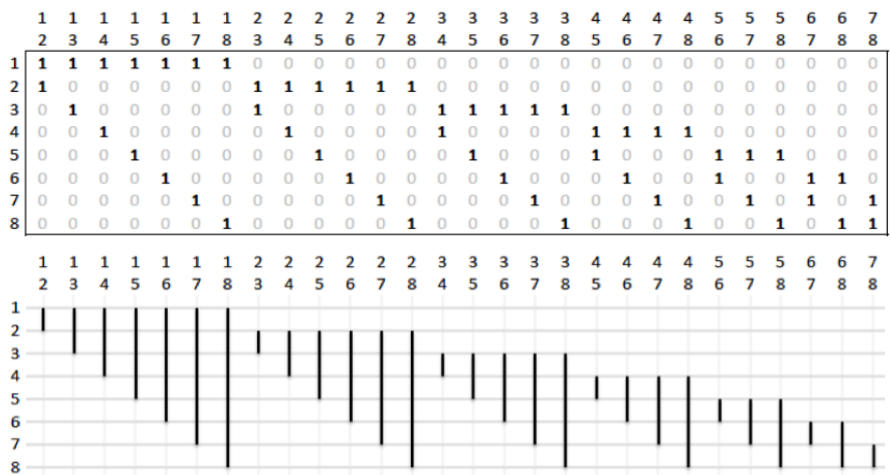

**Fig. 4.** Augmented incidence matrix (top) and cartographic representation (bottom) of $K8$

Now replace the columns of the augmented incidence matrix of $K8$ that represent edges not present in $G$ with all-zero columns; the resulting matrix is the augmented incidence matrix $D$ of the graph $G$. In this matrix, draw black vertical line segments connecting the "1" entries in each column of the augmented incidence matrix, and then replace the matrix with gray horizontal lines at the vertical center of each row and gray line segments at the horizontal center of each column. The result is a cartographic drawing of $G$, as shown in Figure 5:

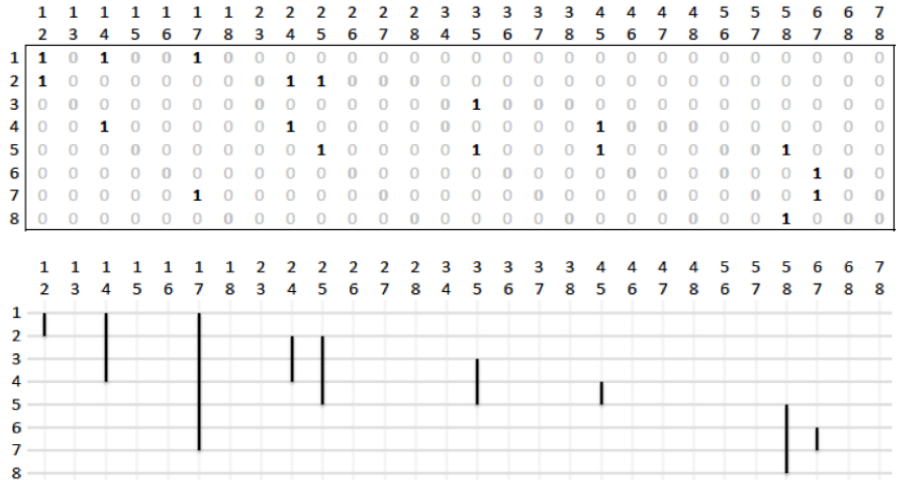

**Fig. 5.** Augmented incidence matrix (top) and cartographic representation (bottom) of $G$

The sources [TU83], [YL87], [MC99], and [LO12] all describe variants of this cartographic representation. The specifics of the representations in these sources vary, but (except for a 90 degree rotation in part of [MC99]) they tend to agree on six basic principles which define a cartographic representation of a graph. We summarize those basic principles below as an aid to readers and, especially, to those who are setting out to produce cartographic drawings of graphs for the first time.

1. A latitude (drawn as a horizontal band) represents a vertex of a graph.

- It can be drawn thin (like the earth's equator).
- It can be drawn thick (like the earth's tropics).
- Various segments of it may be omitted from a drawing.
- It cannot share an $x$ coordinate with, or touch, another latitude.
- There is no need to show or identify its right end or its left end. If these ends exist, they may coincide.
- Descriptions of the vertex it represents may be written on it, or in its left margin, or in its right margin.

2. A longitude (drawn as a bounded segment of a vertical band) represents an edge of a graph.

- It can be drawn thin (like the segment of the international dateline lying in the antarctic circle).
- It can be drawn thick (like the western tenth of Kansas, stretching from North 37 degrees to North 40 degrees).
- No part of it may be omitted from a drawing.
- It cannot share a $y$ coordinate with, or touch, another longitude.
- Its top end must be in the latitude of a vertex. Its bottom end must be in the latitude of a vertex. Each must be visible, and clearly identifiable as ends. If both these ends are on the same latitude, this edge represents a loop.
- Descriptions of the edge it represents may be written on a part of it, or in its top margin, or in its bottom margin.

3. Don't waste ink. Especially colored ink. Don't waste space, either - but use space where it clarifies the meaning of the ink you use; space can convey as much information as ink.

4. Your drawings will usually be landscape shape, not portrait. A tatami video screen is a good canvas to draw on.

5. The complement $\bar{G}$ of a graph $G$ with $n$ vertices is another graph with the same $n$ vertices. $\bar{G}$ contains all the edges not present in $G$ and none of the edges present in $G$ [WE96, p. 3]. The complement $\bar{G}$ of a graph $G$ always presents exactly the same information as $G$, but as often as not, a drawing of the complement $\bar{G}$ is easier to understand than a drawing of $G$. So let the situation at hand suggest whether to present a drawing of $G$, a drawing of $\bar{G}$, or both.

6. Remember that a graph $G$ can be drawn cartographically in many different ways - some of which will be far easier to understand than others. So take the time to rearrange the rows and columns of $G$'s incidence matrix in a few purposeful ways. This will provide several alternative cartographic drawings to choose from.

## 5.1 How to Draw a Digraph

To draw a digraph, place an arrowhead, diamond, or other typographically attractive symbol at the point where the vertical line representing the edge intersects the horizontal line representing the destination vertex. Figure 6 shows the large digraph $H$ (whose node-link representation appears in Figures 2 and 3) drawn in cartographic representation.

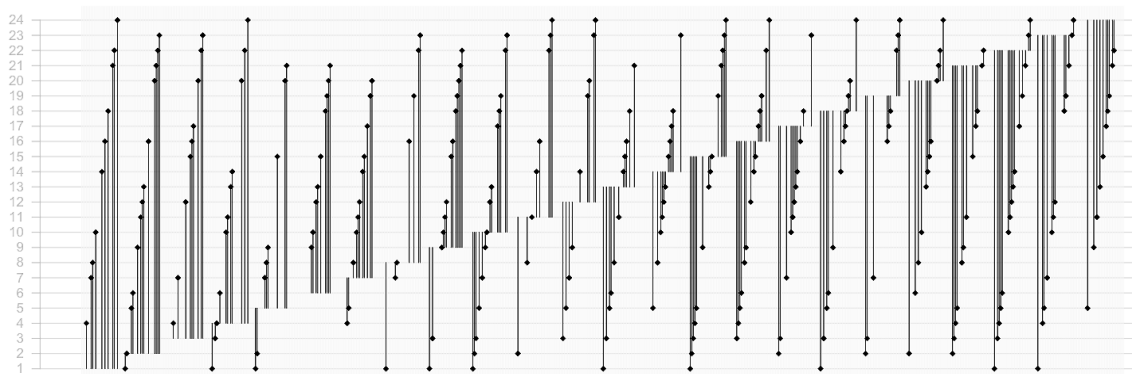

**Fig. 6.** The large digraph $H$ drawn cartographically

Note that in Figure 6 two classes of $x$-coordinate positions have been left blank (that is, unoccupied by a dark vertical line representing an edge):

- Each x coordinate representing an edge which could be present in the graph but is not present (that is, an $x$-coordinate representing an entry in the adjacency matrix whose value is "0") is left blank.

- A few blank positions which do not represent absent edges are left to separate groups of edges representing rows of the adjacency matrix. This is done purely for visual clarity.

### 5.2 Permuting Edge Order

The version of the cartographic representation which draws edges in the order dictated by a row-major traversal of the graph's adjacency matrix (as illustrated in Figure 6) is only one of many possible variations of the cartographic representation.

Many other edge orderings are possible; we divide them into two classes.

**Algorithmic edge orderings.**

Algorithmic edge orderings are generated by running an algorithm which takes the graph's adjacency matrix as input and generates an edge ordering as output. Examples of algorithmic edge orderings of the large graph $H$ are the ordering by origin-vertex out-degree (Figure 7), the ordering by destination-vertex in-degree (Figure 8), and many others not illustrated, such as ordering of a weighted graph by edge weight.

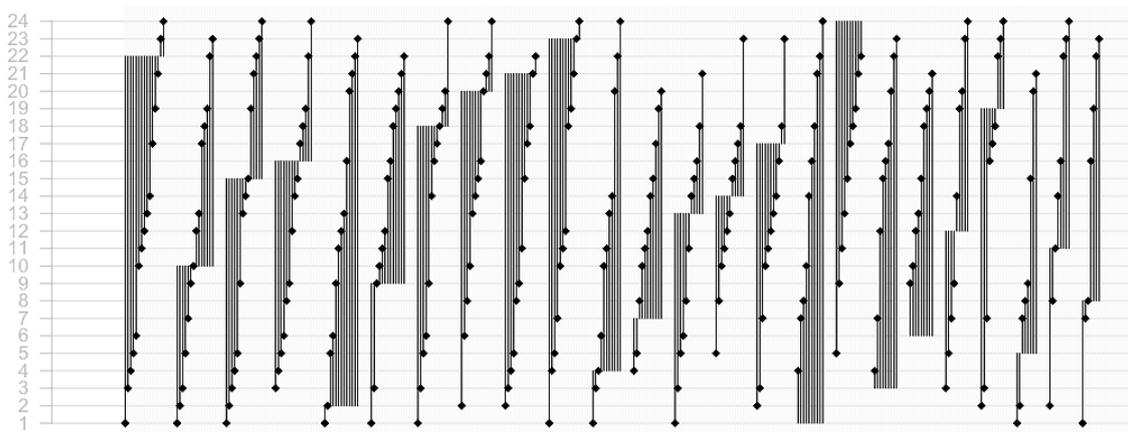

**Fig. 7.** The large digraph $H$ drawn cartographically with edges sorted by origin vertex's out-degree

Note that in Figures 7 and 8, no $x$-coordinates have been left unoccupied to represent absent edges; a few unoccupied $x$-coordinate positions have been left for visual clarity between groups of edges originating from (in Figure 7) or terminating at (in Figure 8) the same vertex.

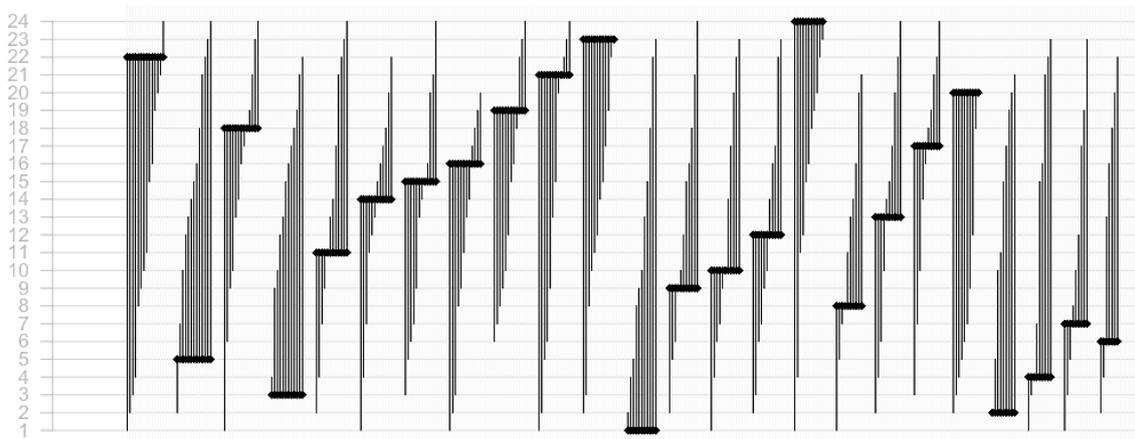

**Fig. 8.** The large digraph $H$ drawn cartographically with edges sorted by destination vertex's in-degree

**Aesthetic edge orderings.**

In some cases it may be desirable to order edges in a way which aids the reader in interpreting some specific semantic aspect of a graph; for example, in a graph whose vertices represent people and whose edges represent telephone calls between people, it might be helpful to criminal investigators to cluster all the edges originating from or terminating at a vertex representing a suspect. We call such an ordering "aesthetic" because its intended effect is to highlight a particular interpretation of the graph. Aesthetic orderings should be constructed with caution to avoid mendacity and obscurantism.

### 5.3   Permuting Vertex Order

Vertex order can also be permuted in cartographic representations of graphs. This will make more sense when vertices are named rather than numbered; vertices can be arranged in any order on the vertical axis. Vertices and edges can be re-ordered independently of one another.

### 5.4   How To Draw a Weighted Graph

A weighted graph can be drawn in several ways using cartographic representation:

(1)   Edges could be colored to indicate their weights - for example, light gray for very small weights, dark gray for small weights, dark red for large weights, and bright red for very large weights.
(2)   Edges could be drawn as in the standard cartographic representation, with annotations below the $x$ axis giving the numeric value of each edge's weight.
(3)   Edges could be drawn using line segments whose widths are linearly or logarithmically proportional to their weights.

# 6 How to Draw Two Graphs

Node-link representations do not lend themselves to comparisons of graphs, or to creating a drawing of the complement of a single graph; cartographic representations make both these activities simple.

## 6.1 How to Illustrate Both a Graph and its Complement With a Single Drawing

A drawing of the complement of a graph can be much more comprehensible than a drawing of the graph itself, particularly for dense graphs. Dense graphs seem to arise frequently in the biological literature.

Figure S6(B) of [CH09], for example, shows seven graphs characterizing acetylation in certain proteins; all seven graphs are nearly complete. [CH09]'s Figure S6(B) consists of seven "clusters", each of which is a graph. The 4th cluster provides a particularly vivid example of the utility of enhancing the comprehensibility of a graph by drawing its complement.

Figure 9 below redraws cluster 4 in the node-link representation; it also shows a node-link representation of cluster 4's complement, and a cartographic representation of cluster 4.

The node-link drawing of cluster 4's complement is much clearer than the node-link drawing of cluster 4 itself; the complement shows clearly that cluster 4 lacks only one edge. But the cartographic representation is even better than node-link drawing of the complement; the single cartographic figure shows at a glance both which edges are present in cluster 4 and which edge is absent.

It is possible, of course, to draw the complement of cluster 4 cartographically; the cartographic drawing has only a single vertical line segment representing the edge connecting vertices 7 and 11.

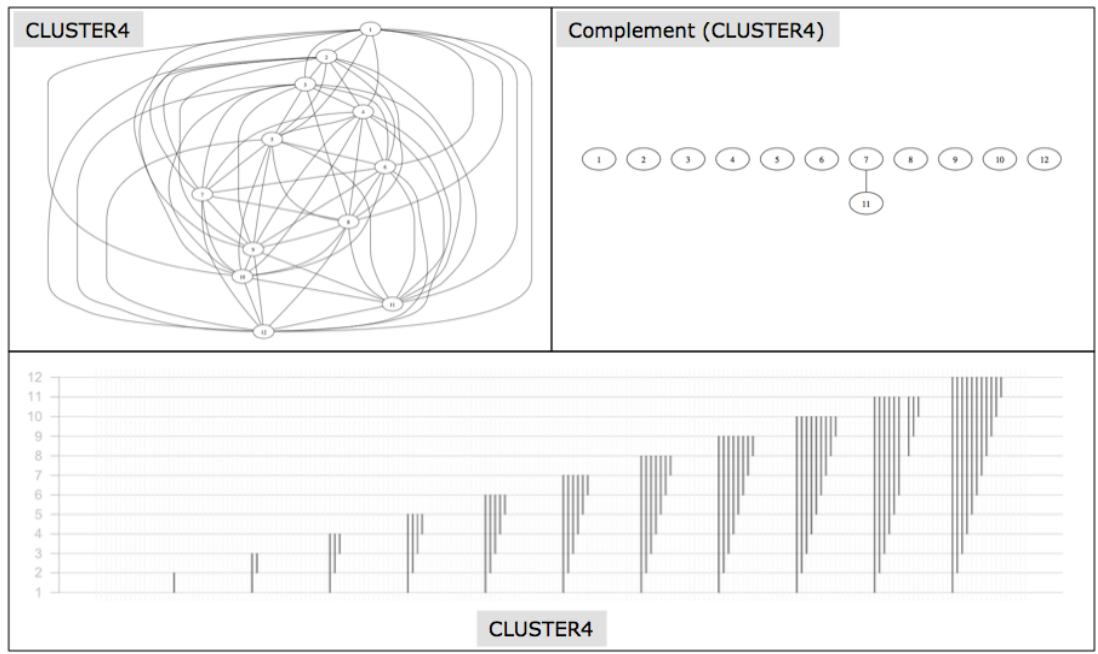

**Fig. 9.** A dense graph and its complement, drawn node-link style (top) and cartographically (bottom)

## 6.2 How to Compare and Contrast Two Graphs With a Single Drawing

The node-link representation provides no support for comparing two similar or related graphs, particularly in cases where the graphs to be compared share some, but not all, vertices. But the need to compare graphs arises frequently in many disciplines. In software engineering, for example, it is often desirable to compare the graph representing the connectivity of modules in a program before a code change with the graph representing module connectivity after the code change. In biology, it is useful to compare the graph of which genes one descendant species inherits from an ancestral species to the graph of which genes another descendant species inherits from the same ancestral species, to determine the degree of relatedness of the two descendant species. And in computer security, it can be useful to compare the graph representing information flow between computers in a network yesterday to the graph representing information flow between computers in the same network today.

The state of the art in making comparisons between two graphs using node-link drawings seems to be to draw each graph individually and place the two drawings side-by-side on the page. In most cases of interest, both graphs' node-link drawings are unintelligible hairballs in their own right. Juxtaposing two unintelligible hairballs merely creates an unintelligible compound hairball. What we really want to know, when we compare two graphs $A$ and $B$ is:

- Which edges are in $A$ but not $B$ (i.e. are in the relative complement $A \setminus B$)?
- Which edges are in $B$ but not $A$ (i.e. are in the relative complement $B \setminus A$)?
- Which edges are in both $A$ and $B$ (i.e. are in the intersection $A \cap B$)?

- Which edges are in either $A$ or $B$ or both (i.e. are in the union $A \cup B$)?
- Which edges are in either $A$ or $B$ but not both (i.e. are in the symmetric difference $A \triangle B$)?

In other words, we want five Boolean operations on pairs of graphs. Oddly enough, most textbooks and standard treatments give only the definition of union [WE96, p. 19] of graphs. This may be because the definition of the union of two graphs has the virtue of depending on the union of their two vertex sets, and because there are no plausible ways to get any worthwhile results from casting the definitions of these other four operations on pairs of graphs in terms of their corresponding operations on pairs of vertex sets. But [WO10] does provide the sought-after definitions of the other four Boolean operations on graph pairs. It turns out that all five Boolean operations on graph pairs $A$ and $B$ should be cast in terms of the union of the vertex sets of $A$ and $B$.

Building a cartographic comparison of two graphs $A = (V, E)$ and $B = (W, F)$ is simple once the Boolean operations are understood. Form the union $D$ of the edge sets $E$ and $F$. Then collect the edges in $D$ which are members only of $E$ to form a set $Q$, collect the edges in $D$ which are members only of $F$ to form a set $S$, and collect all the edges in $D$ which are in neither $Q$ nor $S$ to form a set $R$. Obviously then the graphs $(A \setminus B)$, $(A \cap B)$, and $(B \setminus A)$ are given by

$$A \setminus B = (V \cup W, Q)$$
$$A \cap B = (V \cup W, R)$$
$$B \setminus A = (V \cup W, S)$$

Also,

$$A \cup B = (V \cup W, Q \cup R \cup S)$$
$$A \triangle B = (V \cup W, Q \cup S)$$

All these sets can be represented using a single figure, which we create as follows:

1. Draw all the edges of $(A \setminus B)$ in dark ink on the left-hand side of a cartographic representation.
2. Draw all the edges of $(A \cap B)$ in lighter ink in the center of the same cartographic representation.
3. Draw all the edges of $(B \setminus A)$ in dark ink on the right-hand side of the same cartographic representation.

To illustrate this procedure, and its results, we'll draw the comparison of two attack graphs (graphs used to analyze sequences of actions which can lead to compromise of a system's security) presented in Figures 1 and 2 of [GU07].

The graph $M$ (which appears in Figure 1 of [GU07]) contains 16 vertices and 20 edges. The graph $N$ (which appears in Figure 2 of [GU07]) contains 15 vertices and 17 edges. All the vertices of $N$ also appear in $M$; vertex number 9 of $M$ is missing from $N$. Following the drawing procedure outlined above results in the cartographic drawing illustrated in Figure 10.

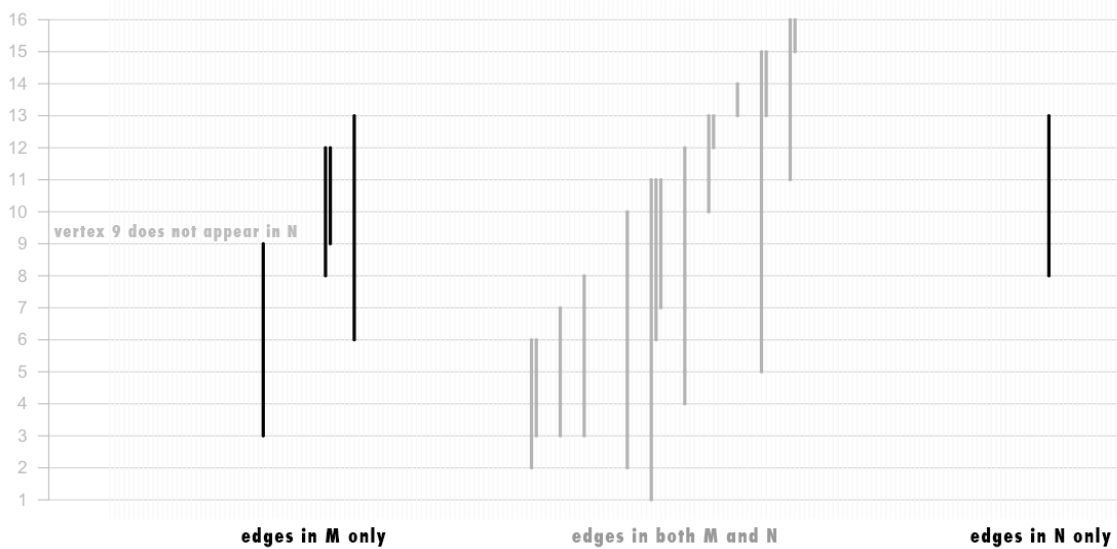

**Fig. 10.** Two graphs compared in a single figure

There are a few things to note about the drawing in Figure 10. First, vertex 9, which doesn't appear in $N$, has been annotated. Second, while the groupings and shadings of the edges in the drawing represent $(M \setminus N)$, $(M \cap N)$, and $(N \setminus M)$ explicitly, the union $(M \cup N)$ and symmetric difference $(M \triangle N)$ are also present, though perhaps less manifestly so. If the lighter-gray edges in the center of the figure are ignored, the collection of black edges on the right and left, taken as a whole, represent the symmetric difference $(M \triangle N)$. And if the differences in darkness of the edges are ignored, the collection of all the edges in the figure represents the union $(M \cup N)$.

## 7 Conclusion

Cartographic drawings of a graph are understandable when node-link drawings aren't, because latitudes are perpendicular to longitudes. No small angles between two bands. No tangential contact. No ambiguity about whether this thing is a vertex or not.

Cartographic approaches to graph displays exalt the incidence matrix. It becomes both a prefiguration and a design tool of a graph drawing. They also shore up the correctness of the Minard/Playfair/Tukey/Tufte [TU83] preference for landscape, rather than portrait, shape. Complete graphs are $n$ high by $n(n-1)/2$ wide. Moreover, cartographic diagrams can say "no", loud and clear. Agree on a universe in advance, and you get absolute complements [WE96] to work with -- a tool logicians and probabilists have handled masterfully for centuries.

Longabaugh's "useful visualization of a network with $10^5$ or even $10^6$ edges" may seem startling to some people accustomed to thinking in terms of node-link drawings of graphs. But his zoom references actually understate his capabilities. The BioFabric software could actually draw a graph that size so that a viewer could see the whole thing synoptically! At 0.5 mm resolution (and Longabaugh often achieves better resolution than that) the

450-vertex *K*450, with 101,025 edges, gives rise to a landscape-oriented cartographic drawing about 0.23 m high and about 50 m wide. Such a drawing can be displayed statically, without requiring zooming, on five turns of a helical vertical ribbon on the inside of a cylinder 4 meters across and less than 2 meters high: a single artwork in a diminutive Guggenheim gallery smaller than some portable swimming pools.

## References

[CH09] Choudhary, C., C. Kumar, F. Gnad, M. Nielsen, M. Rehman, T. C. Walther, J. V. Olsen and M. Mann, Lysine acetylation targets protein complexes and co-regulates major cellular functions, Science, vol. 325, no. 5942, (2009), pp. 834 - 840. (2009)

[GU07] Gupta, S. and J. Winstead, Using attack graphs to design systems, IEEE Security and Privacy, vol. 5, no. 4, July/August (2007), pp. 80 - 83 (2007)

[GY04] Gross, J. and J. Yellen, Handbook of Graph Theory, CRC Press (2004)

[LO12] Longabaugh, W., Combing the hairball with BioFabric: a new approach for visualization of large networks, BMC Bioinformatics (2012) 13: 275 (2012)

[MA13] Ma, K. and C. W. Muelder, Large-scale graph visualization and analytics, IEEE Computer, Volume 46, Number 7, July 2013, pages 39 - 46 (2013)

[MC99] McAllister, A. J., A new heuristic algorithm for the linear arrangement problem, Technical Report 99_126a, University of New Brunswick Faculty of Computer Science (1999)

[TU83] Tufte, E., The Visual Display of Quantitative Information, Cheshire CT: Graphics Press (1983)

[WE96] West, D. B., Introduction to Graph Theory, Prentice-Hall, Upper Saddle River, NJ (1996)

[WO10] Wolfram Research, Inc., Mathematica, Version 8.0, Champaign, IL (2010)

[YL87] Blakley, B., Reduction of Flow Diagrams to Unfolded Form Modulo Snarls, YLYK, Ltd. 14 April 1987 Final Report to AFOSR on Contract F49620-86-C-0103, Defense Technical Information Center (1987)